\documentclass[a4paper,12pt]{article}

\usepackage{mathptm}
\usepackage{setspace}
\usepackage{color}
\usepackage{graphicx}

\usepackage{pdfcolmk}   
\usepackage{multicol}
\usepackage{parentheses}

\usepackage[varg]{txfonts}

 \large\normalsize	

\def\url#1{\textcolor{blue}{\underline{#1}}}	

\begin{document}



\title{Macroscopic models of local field potentials \\ and the
apparent 1/f noise in brain activity}

\author{Claude B\'edard and Alain Destexhe\footnote{Corresponding
author.  Tel: +33 1 69 82 34 35, Fax: +33 1 69 82 34 35, Email:
Destexhe@unic.cnrs-gif.fr} \\ \ \\ Integrative and Computational
Neuroscience Unit (UNIC), \\ CNRS, \\ 1 Avenue de la Terrasse (Bat
33), \\ 91198 Gif-sur-Yvette, France}

\maketitle


\begin{abstract}

The power spectrum of local field potentials (LFPs) has been reported
to scale as the inverse of the frequency, but the origin of this
``$1/f$ noise'' is at present unclear.  Macroscopic measurements in
cortical tissue demonstrated that electric conductivity (as well as
permittivity) is frequency dependent, while other measurements failed
to evidence any dependence on frequency.  In the present paper, we
propose a model of the genesis of LFPs which accounts for the above
data and contradictions. Starting from first principles (Maxwell
equations), we introduce a macroscopic formalism in which macroscopic
measurements are naturally incorporated, and also examine different
physical causes for the frequency dependence.  We suggest that ionic
diffusion primes over electric field effects, and is responsible for
the frequency dependence.  This explains the contradictory
observations, and also reproduces the $1/f$ power spectral structure
of LFPs, {as well as more complex frequency scaling}.  Finally,
we suggest a measurement method to reveal the frequency dependence of
current propagation in biological tissue, and which could be used to
directly test the predictions of the present formalism.

\end{abstract}


\clearpage
\section{Introduction}

Macroscopic measurements of brain activity, such as the
electroencephalogram (EEG), magnetoencephalogram or local field
potentials (LFP), display approximately $1/f$ frequency scaling in
their power spectra~\cite{Bha2001, Bed2006a,Nov1997,Prit1992}.  The
origin of such $1/f$ ``noise'' is at present unclear.  $1/f$ spectra
can result from self-organized critical phenomena~\cite{Jensen1998},
suggesting that neuronal activity may be working according to such
states~\cite{Beggs2003}.  Alternatively, the $1/f$ scaling may be due
to filtering properties of the currents through extracellular
media~\cite{Bed2006a}.  The latter hypothesis, however, was resting
on indirect evidence, and still needs to be examined theoretically,
which is one of the motivations of the present paper.

A {\it continuum model}~\cite{Bed2004} of LFPs incorporated the
inhomogeneities of the extracellular medium into continuous spatial
variations of conductivity ($\sigma$) and permittivity ($\epsilon$)
parameters.  This model reproduced a form of low-pass frequency
filtering in some conditions, while considering the extracellular
medium as locally neutral with $\sigma$ and $\epsilon$ parameters
independent of frequency.  This model was not entirely correct,
however, because macroscopic measurements in cortex revealed a
frequency dependence of electrical parameters~\cite{Gabriel1996}.  We
will show here that it is possible to keep the same model structure
by including plausible causes for the frequency dependence.

In a {\it polarization model}~\cite{Bed2006} of LFPs, the
variations of conductivity and permittivity were considered by
explicitly taking into account the presence of various cellular
processes in the extracellular space around the current source.  In
particular, it was found that the phenomenon of surface
polarization was fundamental to explain the frequency dependence of
LFPs.  The continuum model~\cite{Bed2004} incorporated this effect
phenomenologically through continuous variations of $\sigma$ and
$\epsilon$.  In the polarization model, the extracellular medium is
``reactive'' in the sense that it reacts to the electric field by
polarization effects.  It is also locally non-neutral, which
enables one to take into account the non-instantaneous character of
polarization, which is at the origin of frequency dependence
according to this model~\cite{Bed2006}.  

In the present paper, we propose a ``diffusion-polarization'' model
that synthesizes these previous approaches and which takes into
account both microscopic and macroscopic measurements.  This model
includes ionic diffusion, which we will show has a determinant
influence on frequency filtering properties.  The model also includes
electric polarization, which also influences the frequency-dependent
electric properties of the tissue.  We show that taking into account
ionic diffusion and electric polarization allows us to quantitatively
account for the macroscopic measurements of electric conductivity in
cortical tissue according to the experiments of Gabriel et
al.~\cite{Gabriel1996}.

However, recent measurements of Logothetis et al.~\cite{Logo2007}
evidenced that the frequency dependence of cortical tissue was
negligible, therefore in contradiction with the measurements of
Gabriel et al.~\cite{Gabriel1996}.  We show here that the
``diffusion-polarization'' model can be consistent with both types of
experiments, and thus may reconcile this contradiction.  We will also
examine whether this model can also explain the $1/f$ frequency
scaling observed in LFP or EEG power spectra.  Finally, we consider
possible ways for experimental test of the predictions of this model.

The final goal of this approach is to obtain a model of local field
potentials which is consistent with both macroscopic measurements of
conductivity and permittivity, and the microscopic features of the
structure of the extracellular space around the current sources. 


\section{Materials and Methods}

The numerical simulations presented in Section~\ref{sec-sim} describe 
the impedance of the extracellular medium based on the following 
equation
\begin{equation}
Z(r_1,\omega)=\frac{1}{4\pi}
\int_{r_1}^{\infty}dr'\frac{1}{r'^2}~\frac{1}{\sigma^M(r',\omega) + i\omega\epsilon^M(r',\omega)} ~ .
\label{eqnum1}
\end{equation}
This equation gives the $\omega$-frequency component of the impedance
at point $r_1$ in extracellular space, in spherical symmetry (see
ref.~\cite{Bed2004} and Eq.~\ref{eqgeneralM} for details).

To evaluate this equation, we use MATLAB, which computes the 
Riemann sum :
\begin{equation}
Z(r_1,\omega)=\frac{1}{4\pi}\sum_{r_1}^{N}\frac{\Delta r'}{r'^2}~\frac{1}{\sigma^M(r',\omega) + i\omega\epsilon^M(r',\omega)} ~ .
\label{eqnum2}
\end{equation}
where $\Delta r'$ is the integration step ($ 1~ \mu m$) and $N$ is
determined for a ``slice'' of 1~mm.

We also use the parametric model of Gabriel et al.~\cite{Gabriel1996}
to simulate the frequency dependence of electrical parameters
$\sigma$ and $\epsilon$ of the extracellular fluid from gray matter
(at a temperature of $37^\circ C$).  This model is valid for
frequencies included in the range of 10~Hz to $4\times
10^8$~Hz~\cite{Gabriel1996c}.  According to this model, the absolute
complex and macroscopic permittivity and conductivity (measured
between 10 and $10^{10}$~Hz) in cortical gray matter is 
given by the Cole-Cole parametric model~\cite{Cole-Cole1941}:
\begin{equation}
\epsilon_{\omega}^{*}=-i\frac{\sigma^*}{\omega} = \epsilon_{\infty}\epsilon_o + \epsilon_o\sum_{n=1}^{n=4} \frac{\Delta\epsilon_n}{1+(i\omega\tau_n)^{1-\alpha_n}} -i\frac{\sigma}{\omega}~,
\label{Eq4}
\end{equation}
where the sum runs over 4 polarization modes $n$, $\epsilon_o =
8.85\times 10^{-12}~F/m$ is vacuum permittivity, $\epsilon_{\infty}
= 4.0$ is the permittivity relative to $f=10^{10}~ Hz$, 
$\sigma =0.02~ S/m$ is the static conductivity at $f=0~ Hz$ according 
to the chosen parametric model.  The parameters under the sum of
Eq.~\ref{Eq4} are given in Table~1.

\ \\ \centerline{------------------------ Table~1 here
------------------------}

\section{Results}

We start by outlining a macroscopic model with frequency-dependent
electrical parameters (Section~\ref{sec-theor}), and we discuss the
main physical causes for this frequency dependence
(Section~\ref{sec-freqdep}).  We then constrain the model to
macroscopic measurements of electrical parameters, and provide
numerical simulations to test the model and reproduce the
experimental observations (Section~\ref{sec-sim}).  Finally, we
propose a possible way to test the model experimentally
(Section~\ref{sec-test}).

\subsection{Macroscopic model of local field potentials}
\label{sec-theor}

In this section, we derive the equations governing the time evolution
of the extracellular potential.  We follow a formalism similar to the
one developed previously~\cite{Bed2004}, except that we
reformulate the model macroscopically, to allow the electrical
parameters (the conductivity $\sigma$ and permittivity $\epsilon$) to
depend on frequency, as demonstrated by macroscopic
measurements~\cite{Gabriel1996a,Gabriel1996,Gabriel1996c}.  The
physical causes of this macroscopic frequency dependence will
be examined in Section~\ref{sec-freqdep}.

\subsubsection{General formalism}

We begin by deriving a general equation for the electrical potential
when the electrical parameters are frequency dependent.  We start
from Maxwell equations, taking the 1$^{st}$ and the divergence of the
4$^{th}$ Maxwell equation in a medium with constant magnetic
permeability, giving:
\begin{equation}\label{eqtemps}
\begin{array}{c}
\nabla\cdot\vec{D} = \rho^{free}\\ \\
\nabla\cdot\vec{j} + \frac{\partial \rho^{free}}{\partial t} = 0~.
\end{array}
\end{equation}
where $\vec{D}$, $\vec{j}$ and $\rho^{free}$ are respectively the 
electric displacement, current density and charge density in the
medium surrounding the sources.

Moreover, in a linear medium the equations linking the electric field
$\vec{E}$ with electric displacement $\vec{D}$, and with current
density $\vec{j}$, gives:
\begin{equation}
\vec{D}(\vec{x},~t) = \int_{-\infty}^{\infty} \epsilon 
  (\vec{x},~\tau)\vec{E}(\vec{x},~t-\tau) d\tau \label{link1} \\ \\
\end{equation}
and
\begin{equation}
\vec{j}(\vec{x},~t) = \int_{-\infty}^{\infty} \sigma 
  (\vec{x},~\tau )\vec{E}(\vec{x},~t-\tau)d\tau ~.  \label{link2}
\end{equation}

The Fourier transforms of these equations are respectively
$\vec{D_{\omega}} = \epsilon_{\omega} \vec{E_{\omega}}$ and
$\vec{j_{\omega}} = \sigma_{\omega} \vec{E_{\omega}}$, where we allow
$\sigma$ and $\epsilon$ to depend on frequency.

Given the limited precision of measurements, we can consider
$\nabla\times \vec{E}\approx 0$ for frequencies smaller than 1000~Hz.
Thus, we can assume that $\vec{E} = -\nabla V$ such that the 
complex Fourier transform of Eqs.~\ref{eqtemps} can be written as:
$$
\begin{array}{c}
\nabla\cdot (\epsilon_{\omega}(\vec{x})\nabla V_{\omega}) = - \rho_{\omega}^{free}
\nonumber
\\\nonumber \\\nonumber
 \nabla\cdot (\sigma_{\omega}(\vec{x} )\nabla V_{\omega}) = i\omega\rho_{\omega}^{free}~
\nonumber
\end{array}
$$
Consequently, we have
\begin{equation}\label{eqgeneral}
\nabla\cdot((\sigma_{\omega}+i\omega\epsilon_{\omega})\nabla V_{\omega}) =
\nabla(\sigma_{\omega}+i\omega\epsilon_{\omega})
\cdot\nabla V_{\omega}
+\cdot(\sigma_{\omega}+i\omega\epsilon_{\omega})\nabla^2 V_{\omega}=
0
\end{equation}

Compared to previous derivations (see Eq.~49 in
ref.~\cite{Bed2004}), this equation is a more general form in which
the electrical parameters can be dependent on frequency.

\subsubsection{Macroscopic model}

In principle, it is sufficient to solve Eq.~\ref{eqgeneral} in the
extracellular medium to obtain the frequency dependence of LFPs. 
However, in practice, this equation cannot be solved because the
structure of the medium is too complex to properly define the limit
conditions.  The associated values of electric parameters must be
specified for every point of space and for each frequency, which
represents a considerable difficulty.  One way to solve this problem
is to consider a macroscopic or mean-field approach.  This approach
is justified here by the fact that the values measured experimentally
are averaged values, which precision depends on the measurement
technique.  Because our goal is to simulate those measured values, we
will use a macroscopic model, in which we take spatial averages of
Eq.~\ref{eqgeneral}, and make a continuous approximation for the
spatial variations of these average values. {This type of
approximation can be found in the classic theory of electromagnetism
(see for example Chapters 9 and 10 in Maxwell's treaty for
electricity and magnetism~\cite{Maxwell1873})}.

To this end, we define macroscopic electric parameters, $\epsilon^M$
and $\sigma^M$, as follows:
$$
\epsilon_{\omega}^M(\vec{x}) = <\epsilon_{\omega}(\vec{x})>_{|_V}=f(\vec{x},\omega)
$$
and
$$
\sigma_{\omega}^M(\vec{x}) = <\sigma_{\omega}(\vec{x})>_{|_V}=g(\vec{x},\omega) ~ .
$$
where $V$ is the volume over which the spatial average is taken.  We
assume that $V$ is of the order of $\mu m^3$, and is thus much
smaller than the cortical volume, so that the mean values will be
dependent of the position in cortex.

Because the average values of electric parameters are statistically
independent of the mean value of the electric field, we have:
$$
<\vec{j}^{total}>_{|_V}(\vec{x},~t) = \int_{-\infty}^{\infty}\sigma^M(\tau)
<\vec{E}>_{|_V}(\vec{x},~t-\tau)d\tau +
\int_{-\infty}^{\infty}\epsilon^M(\tau)\frac{\partial
<\vec{E}>_{|_V}}{\partial t}(\vec{x},~t-\tau ) d\tau ~ ,
$$ 
where the first term in the righthand represents the
``dissipative'' contribution, and the second term represents the
``reactive'' contribution (reaction from the medium).  Here, all
physical effects, such as diffusion, resistive and capacitive
phenomena, are integrated into the frequency dependence of
$\sigma^M$ and $\epsilon^M$. We will examine this frequency
dependence more quantitatively in Section~\ref{sec-freqdep}.

The complex Fourier transform of $<\vec{j}^{total}>_{|_V}(\vec{x},~t)$ then 
becomes:
\begin{equation}
<\vec{j}_{\omega}^{total}>_{|_V}= (\sigma_{\omega}^M
+i\omega\epsilon_{\omega}^M)<\vec{E}_{\omega}>_{|_V}) =
\sigma_{z}^{M}<\vec{E}_{\omega}>_{|_V} ~ ,
\label{dcou}
\end{equation}
where $\sigma_{z}^{M}$ is the complex conductivity.  We can also
assume
\begin{equation}
\sigma_{z}^M = i\omega\epsilon_{z}^{M}
\end{equation}
such that
\begin{equation}\label{ff}
\nabla\cdot<\vec{j}_{\omega}^{total}>_{|_V}=\nabla\cdot
(\sigma_{z}^M<\vec{E_{\omega}>_{|_V}}) =\nabla\cdot
(i\omega\epsilon_{z}^M<\vec{E_{\omega}}>_{|_V})=0 ~ .
\label{comp}
\end{equation}

Because $\sigma_{z}^M=(\sigma_{\omega}^M
+i\omega\epsilon_{\omega}^M)$ and $\vec{E_{\omega}}>=-\nabla
<V_{\omega}>$, the expressions above (Eqs.~\ref{ff}) can also be
written in the form: 
\begin{equation}\label{eqgeneralM}
  \nabla\cdot((\sigma_{\omega}^M+i\omega\epsilon_{\omega}^M)\nabla
  <V_{\omega}>_{|_V}) =0 ~ .
\end{equation}

We note that starting from the continuum model~\cite{Bed2004}, where
only spatial variations were considered, and generalizing this model
by including frequency-dependent electric parameters, gives the same
mathematical form as the original model (compare with Eq~49 in
ref.~\cite{Bed2004}).  This form invariance will allow us to
introduce in Section~\ref{sec-freqdep} the surface polarization
phenomena by including an {\it ad hoc} frequency dependence in
$\sigma_{\omega}^M$ and $\epsilon_{\omega}^M$.  The physical causes of this macroscopic
frequency dependence is that the cortical medium is microscopically
non-neutral (although the cortical tissue is macroscopically
neutral).  Such a local non-neutrality was already postulated in a
previous model of surface polarization~\cite{Bed2006}.  This
situation cannot be accounted by Eq.~\ref{eqgeneral} if $\sigma_{\omega}^M$
and $\epsilon_{\omega}^M$ are frequency independent (in which case
$\rho_{\omega} =0$ when $\nabla V_{\omega} =0$). Thus,
including the frequency dependence of these parameters enables the
model to capture a much broader range of physical phenomena.

Finally, a fundamental point is that the frequency dependences of
the electrical parameters $\sigma_{\omega}^M$ and $\epsilon_{\omega}^M$ cannot take
arbitrary values, but are related to each-other by the
Kramers-Kronig relations~\cite{Foster1989,Kronig1926,Landau1981}: 
\begin{equation}
\Delta\epsilon^M(\omega) =  \frac{2}{\pi} \fint_{0}^{\infty}\frac{~\sigma^M(\omega')}{\omega'^2-\omega^2}d\omega' 
\label{kronig1}
\end{equation}
and
\begin{equation}
\sigma^M(\omega) =\sigma^M(0) - \frac{2\omega^2}{\pi} \fint_{0}^{\infty}\frac{\Delta \epsilon^M(\omega')}{\omega'^2-\omega^2}d\omega' 
\label{kronig2}
\end{equation}
where principal value integrals are used.  These equations are
valid for any linear medium (i.e., when Eqs.~\ref{link1} and
\ref{link2} are linear).  These relations will turn out to be
critical to relate the model to experiments, as we will see below.

Note that, contrary to frequency dependence, the spatial dependences
of $\sigma_{\omega}^M$ and $\epsilon_{\omega}^M$ are independent of each other,
because these dependences are related to the spatial distribution
of elements within the extracellular medium.

\subsubsection{Simplified geometry for macroscopic parameters}
\label{simplif}

To obtain an expression for the extracellular potential, we need to
solve Eq.~\ref{eqgeneralM}, which is possible analytically only if
we consider a simplified geometry of the source and surrounding
medium.  The first simplification is to consider the source as
monopolar.  The choice of a monopolar source does not intrinsically
reduce the validity of the results because multipolar
configurations can be composed from the arrangement of a finite
number of monopoles~\cite{Purc1984}.  In particular, if the
physical nature of the extracellular medium determines a frequency
dependence for a monopolar source, it will also do so for
multipolar configurations.  A second simplification will be to
consider that the current source is spherical and that the
potential is uniform on its surface.  This simplification will
enable us to calculate exact expressions for the extracellular
potential and should not affect the results on frequency
dependence.  A third simplification is to consider the
extracellular medium as isotropic.  This assumption is certainly
valid within a macroscopic approach, and justified by the fact that
the neuropil of cerebral cortex is made of a quasi-random
arrangement of cellular processes of very diverse
size~\cite{Braitenberg}.  This simplified geometry will allow us to
determine how the physical nature of the extracellular medium can
determine a frequency dependence of the LFPs, independently of
other factors (such as more realistic geometry, propagating
potentials along dendrites, etc).

Thus, considering a spherical source embedded in an isotropic
medium with frequency-dependent electrical parameters, combining
with Eq.~\ref{eqgeneralM}, we have: 
\begin{equation}
\frac{d^{2}<V_\omega>_{|_V}}{dr^{2}} +
\frac{2}{r} \frac{d<V_\omega>_{|_V}}{dr} +
\frac{1}{(\sigma_{\omega}+i\omega\epsilon_{\omega})}
\frac{d(\sigma_{\omega}+i\omega\epsilon_{\omega})}{dr} \frac{d<V_\omega>_{|_V}}{dr} = 0 ~ .
\end{equation}
Integrating this equation gives the following relation between two
points $r_1$ and $r_2$ in the extracellular space,
\begin{equation}
 r_1^{2} \ \frac{d<V_\omega>_{|_V}}{dr}(r_1) \
\left[ \sigma_{\omega}(r_1) + i \omega \epsilon_{\omega}(r_1) \right] = r_2^{2} \
\frac{d<V_\omega>_{|_V}}{dr}(r_2) \ \left[ \sigma_{\omega}(r_2) + i \omega
\epsilon_{\omega}(r_2) \right] ~ . 
\end{equation}
Assuming that the extracellular potential vanishes at large distances
($<V_\omega>=0$), we find
\begin{equation}
<V_\omega>_{|_V}(r_1) = \frac{I_\omega}{4 \pi
} \int_{r_1}^{\infty} dr' \ \frac{1}{r'^{2}} \
\frac{1}{\sigma_{\omega}(r') + i
\omega~\epsilon_{\omega}(r')} ~ . 
\end{equation}

This equation is analogous to a similar expression derived
previously (Eq.~25 in ref.~\cite{Bed2004}), but more general.  The
two formalisms are related by:
$$
<\vec{j}_{\omega}>_{|_V}=\sigma_z<\vec{E}_{\omega}>_{|_V}=
(\sigma_{\omega}^M+i\epsilon_{\omega}^M)<\vec{E}_{\omega}>_{|_V}
$$
instead of $\vec{j_{\omega}}= \sigma^M\vec{E}_{\omega}$ (see Eq.~4 in
ref.~\cite{Bed2004}).  This difference is due to the fact that the
conductivity here depends on frequency.

In the following, we will use the simplified notations
$\vec{j}_{\omega}$,$\vec{E}_{\omega}$ and $V_{\omega}$ instead of
$<\vec{j_{\omega}}>_{|_V}$,$<\vec{E_{\omega}}>_{|_V}$ and
$<V_{\omega}>_{|_V}$, respectively.

Using the relation $V_{\omega} = Z_{\omega} I_{\omega}$, the impedance
 $Z_{\omega}$ is given by:
\begin{equation}
Z_{\omega}(r)=\frac{1}{4\pi}
\int_{r}^{\infty}dr'\frac{1}{r'^2}~\frac{1}{\sigma_{\omega}^M(r')[1 + i
\omega \tau_{\omega}(r')]}
\label{impedance}
\end{equation}
where $\tau_{\omega}(r') =
\frac{\epsilon_{\omega}^M}{\sigma_{\omega}^M}$ and $r$ is the
distance between the center of the source and the position defined by
$\vec{r}$.

\subsection{Physical causes for frequency-dependent electrical
parameters} \label{sec-freqdep}

In the following, we successively consider two different cases
of extracellular medium: first, ``non-reactive'' media, in which the
current passively flows into the medium; second, ``reactive'' media,
in which some properties (such as charge distribution) may change
following current flow.  For each medium, we will consider two types
of physical phenomena, the current produced by the electric field,
and the current produced by ionic diffusion, {as schematized in
Fig.~\ref{polariz}).}

\subsubsection{Non-reactive media with electric fields (Model N)}

``Non reactive'' media ($\frac{\omega
\epsilon_{\omega}^M}{\sigma_{\omega}^M}<<1, \sigma_{\omega}^M
=\sigma^M$ and $\epsilon_{\omega}^M=\epsilon^M$) are equivalent to
``resistive'' media, in which the resistance (or equivalently, the
conductivity) does not change following the flow of current.  The
simplest type of such configuration consists of a resistive medium
(such as a homogeneous conductive fluid) in which current sources
solely interact via their electric field.  Applying
Eq.\ref{impedance} to this configuration is equivalent to model the
extracellular potential by Coulomb's law:
\begin{equation}
  V_{\omega}(\vec{r}) = \frac{1}{4 \pi \sigma^M}\cdot \frac{I_{\omega}}{r} ~ ,
\end{equation}
where $V_{\omega}(\vec{r})$ is the extracellular potential at a
position defined by $\vec{r}$ in extracellular space, $r$ is the
absolute distance between $\vec{r}$ and the center of the current
source.  Here, the conductivity ($\sigma_{\omega}^M(r)=\sigma^M$) is
independent of space and frequency, and thus, this model is not
compatible with macroscopic measurements of frequency
dependence~\cite{Gabriel1996a,Gabriel1996,Gabriel1996c}.  It is,
however, the most frequently used model to calculate extracellular
field potentials~\cite{Nu2006}.  This model will be referred to as
``Model N'' in the following.

\subsubsection{Non-reactive media with ionic diffusion (Model D)}

Because current sources are ionic currents, there is flow of ions
inside or outside of the membrane, and another physical phenomena
underlying current flow is ionic diffusion.  Let us consider a
resistive medium such as a homogeneous extracellular conductive fluid
with electric parameters 
$$
\sigma_z^m=\sigma_{\omega}^m(r)(1 +i\frac{\omega\epsilon_{\omega}^m(r)}{\sigma_{\omega}^m(r)})=\sigma^m(1 +i\frac{\omega\epsilon^m}{\sigma^m})\approx \sigma^m~,
$$
and in which the ionic diffusion coefficient is $D$.  When the
extracellular current is exclusively due to ionic diffusion, the
current density depends on frequency as $\sqrt{\omega}$ (see
Appendix~\ref{secscal}).  A resistive medium behaves as if it had a
resistivity equal to $a(1 + \frac{b}{\sqrt{\omega}})$, where $b$ is
complex.  The parameter $a$ is the resistivity for very high
frequencies, and reflects the fact that the effect of ionic diffusion
becomes negligible compared to calorific dissipation (Ohm's law) for
very high frequencies.  When ionic diffusion is dominant compared to
electric field effects, the real part of $b$ is much larger than $a$.

The frequency dependence of conductivity will be given by: 
\begin{equation}
 \sigma_{\omega}^M=\frac{\sigma^{m}\sqrt{\omega~ }}{\sqrt{\omega~} +k}
\label{diffus}
\end{equation}

Applying Eq.~\ref{impedance} to this configuration gives the
following expression for the electric potential as a function of
distance:
\begin{equation}
  V_{\omega}(\vec{r}) = 
\frac{1}{4 \pi \sigma_{z}^M} \cdot\frac{I_{\omega}}{r}
=\frac{\sqrt{\omega~} + k}{\sqrt{\omega~}}\cdot\frac{1}{4 \pi\sigma^m} \cdot\frac{I_{\omega}}{r}
\end{equation}

This expression shows that, in a non-reactive medium, when the
extracellular current is dominated by ionic diffusion (compared to
that directly produced by the electric field), then the conductivity
will be frequency dependent and will scale as $\sqrt{\omega}$. 
This model will be referred to as ``Model D'' in
the following.  Note that, if the electric field primes over ionic
diffusion, then we have the situation described by Model~N above.

\subsubsection{Reactive media with electric fields (Model P)}

In reality, extracellular media contain different charge densities,
for example due to the fact that cells have a non-zero membrane
potential by maintaining differences of ionic concentrations between
the inside and outside of the cell.  Such charge densities will
necessarily be influenced by the electric field or by ionic
diffusion.  As above, we first consider the case with only electric
field effects and will consider next the influence of diffusion and
the two phenomena taken together.

Electric polarization is a prominent type of ``reaction'' of the
extracellular medium to the electric field.  In particular, the
ionic charges accumulated over the surface of cells will migrate
and polarize the cell under the action of the electric field.  It
was shown previously in a theoretical study that this ``surface
polarization'' phenomena can have important effects on the
propagation of local field potentials~\cite{Bed2006}.  If a charged
membrane is placed inside an electric field $\vec{E}_0^S$, there is
production of a secondary electric field $\vec{E}_{\omega}^S$ given
by (see Eq.~31 in ref.~\cite{Bed2006}):
\begin{equation}
 \vec{E}_{\omega}^S = \frac{\vec{E}_{0}^S}{1+i\omega\tau_M} ~ .
\label{field}
\end{equation}

This expression is the frequency-domain representation of the effect
of the inertia of charge movement associated with surface
polarization, reflecting the fact that the polarization does not
occur instantaneously but requires a certain time to setup.  This
frequency dependence of the secondary electric field was derived in
ref.~\cite{Bed2006} for a situation where the current was exclusively
produced by electric field.  The parameter $\tau_M$ is the
characteristic time for charge movement (Maxwell-Wagner time) and
equals $\epsilon^{memb} / \sigma^{memb}$, where $\epsilon^{memb}$ and
$\sigma^{memb}$ are respectively the absolute (tangential)
permittivity and conductivity of the membrane surface, respectively,
and are in general very different from the permittivity and the
conductivity of the extracellular fluid.

Let us now determine for zero-frequency the amplitude of the
secondary field $\vec{E}_0^S$ produced between two cells embedded
in a given electric field.  First, we assume that it is always
possible to trace a continuous path which links two arbitrary
points in the extracellular fluid (see Fig.~\ref{ModeleSimple}B). 
Consequently, the domain defined by extracellular fluid is said to
be {\it linearly connex}.  In this case, the electric potential
arising from a current source is necessarily continuous in the
extracellular fluid.  Second, in a first approximation, we can
consider that the cellular processes surrounding sources are
arranged randomly (by opposition to being regularly structured) and
their distribution is therefore approximately isotropic. 
Consequently, the field produced by a given source in such a medium
will also be approximately isotropic.  Also consequent to this
quasi-random arrangement, the equipotential surfaces around a
spherical source will necessarily cut the cellular processes around
the source (Fig.~\ref{ModeleSimple}B).

{Suppose that at time $t=0$, an excess of charge appears at a
given point in extracellular space, then a static electric field is
immediately produced.  At this time, currents begins to flow in
extracellular fluid, as well as inside the different cellular
processes surrounding the source.  These cells begin to polarize,
with $\tau_{MW}$ as the characteristic polarization time. 
Asymptotically, the system will reach an equilibrium where the
polarization will neutralize the electric field, such that there is
no electric field inside the cells (zero current).  Now suppose that,
in the asymptotic regime, there would still be a current flowing in
between cells (in the extracellular fluid), then we have two
possibilities.  First, the equipotential surfaces are discontinuous,
or they cut the membrane surfaces (as illustrated in
Fig.~\ref{ModeleSimple}B).  The first possibility is impossible
because it would imply an infinite electric field.  The second
possibility is also impossible, because cells are isopotential due to
polarization.  Therefore, we can say that, asymptotically, there is
no current flowing in extracellular fluid at $f=0$, and necessarily
this is equivalent to a dielectric medium.  In other words, {\it a
passive inhomogeneous medium with randomly distributed passive cells
is a perfect dielectric at zero frequency}.  In this case, the
conductivity must tend to 0 when frequency tends to 0.  Thus, in the
following, we assume that $\vec{E}_{0}^S=-\vec{E}_{0}^P$ where
$\vec{E}_{0}^P$ is the field produced by the source.}

\ \\ \centerline{------------------------ Figure~\ref{ModeleSimple} here
------------------------}

It follows that the expression for the current density in
extracellular space as a function of the electric field is given by:
$$
 \vec{j}_{\omega} = \sigma_{z}^M\vec{E}_{\omega}^P= \sigma^m~\cdot~
(1+i\frac{\omega\epsilon^m}{\sigma^m})~\cdot~ (\vec{E}^P +\vec{E}^S) =
 \sigma^m\cdot~
(1+i\frac{\omega\epsilon^m}{\sigma^m})~\cdot~ \frac{ i\omega\tau_M}{1+i\omega\tau_M}~\cdot~\vec{E}_{\omega}^P
$$

In addition, for cerebral cortical tissue, we have
$1+\frac{\omega\epsilon^m}{\sigma^m}\approx 1$ for frequencies $>10~
Hz$ and $<1000~Hz$ (see ref.~\cite{Gabriel1996}.  Thus, an excellent
approximation of the conductivity can be written as:
\begin{equation}
 \sigma_{z}^M = \sigma^m\cdot\frac{i\omega\tau_M}{1+i\omega\tau_M}
\end{equation}
Applying Eq.~\ref{impedance} gives:
\begin{equation}
  V_{\omega}(\vec{r}) \ = \
\frac{1}{4 \pi \sigma_{z}^M} \cdot\frac{I_{\omega}}{r} \ = \
\frac{i\omega\tau_M}{1+i\omega\tau_M}\cdot\frac{1}{4 \pi\sigma^m} \cdot\frac{I_{\omega}}{r}
\end{equation}

This model describes the effect of polarization in reaction to the
source electric field, and will be referred to as ``Model P' in the
following.

\subsubsection{Reactive media with electric field and ionic diffusion
(Model DP)}

\ \\ \centerline{------------------------ Figure~\ref{polariz} here
------------------------}

The propagation of current in the medium is dominated by ionic
diffusion currents or by currents produced by the electric field,
according to the values of $k$ and $k_1$ with respect to
$\sqrt{\omega~ }$.  The values of $k$ and $k_1$ are respectively
inversely proportional to the square root of the global ionic
diffusion coefficient in the extracellular fluid, and of membrane
surface (see Appendix~\ref{secscal}).  

We apply the reasoning based on the connex topology of the cortical
medium (see above) to deduce the order of magnitude of the induced
field for zero frequency $\vec{E}_0^S$
\begin{equation}
 \vec{E}_{\omega}^S = -\frac{\vec{E}_{0}^P}{1+i\sqrt{\omega~}\tau} ~ .
\label{field2}
\end{equation}
where
$$
\tau = (\sqrt{\omega~ } +k_1)\tau_M
=\sqrt{\omega }\frac{\epsilon^{memb}}{\sigma^{memb}} ~ .
$$

Because the ``tangential'' conductivity on membrane surface is given
by
$$
 \sigma_{\omega}^{memb}=\frac{\sigma^{Memb}\sqrt{\omega~ }}{\sqrt{\omega~} +k_1}
$$
when the current is dominated by either electric field or ionic
diffusion (see Eq.~\ref{diffus}).

It follows that the expression for the current density in
extracellular space as a function of the electric field is given by
$$
 \vec{j}_{\omega} = \sigma_{z}^M\vec{E}_{\omega}^P= \frac{\sigma^m\sqrt{\omega}}{\sqrt{\omega} +k}~\cdot~
(1+i\frac{\omega\epsilon_{\omega}^m}{\sigma_{\omega}^m})~\cdot~ (\vec{E}^P +\vec{E}^S) \approx
 \frac{\sigma^m\sqrt{\omega}}{\sqrt{\omega} +k}~\cdot~ \frac{ i\sqrt{\omega~}~\tau}{1+i\sqrt{\omega~}~\tau}~\cdot~\vec{E}_{\omega}^P
$$
because
$1+i\frac{\omega\epsilon_{\omega}^m}{\sigma_{\omega}^m}\approx 1$ in
cortical tissue for frequencies $>10~Hz$ and $<1000~Hz$ (see
ref.~\cite{Gabriel1996}).

Thus, we have the following expression for the complex conductivity 
of the extracellular medium:
\begin{equation}
 \sigma_z^M = \sigma_{\omega}^M +i\omega\epsilon_{\omega}^M=
\frac{\sigma^m\sqrt{\omega}}{\sqrt{\omega}
+k}~\cdot~
\frac{ i\sqrt{\omega~}~\tau}{1+i\sqrt{\omega~}~\tau}~\cdot~
\label{apparent}
\end{equation}
where $\tau= (\sqrt{\omega} +k_1)\tau_M$.

Thus, we have obtained a unique expression (Eq.~\ref{apparent}) for
the apparent conductivity in extracellular space outside of the
source, and its frequency dependence due to differential Ohm's law,
electric polarization phenomena and ionic diffusion.  These phenomena
are responsible for an apparent frequency-dependence of the electric
parameters, which will be compared to the frequency dependence
observed in macroscopic measurements of conductivity
(Section~\ref{sec-sim}).

Finally, Eqs.~\ref{impedance} and \ref{apparent} imply that the
macroscopic impedance of a homogeneous spherical shell of width
$R_2-R_1$ is given by:
\begin{equation}
Z_{\omega}=\frac{1}{4\pi}
\int_{R_1}^{R_2}\frac{1}{r'^2}~\frac{dr'}{\sigma_{\omega }^M + i
\omega \epsilon_{\omega }^M}
=\frac{R_2-R_1}{4\pi R_1R_2}~\cdot\frac{1}{\sigma_{\omega }^M + i
\omega \epsilon_{\omega }^M}~.
\label{impedance1}
\end{equation}

In the following, this model will be referred as the
``diffusion-polarization'' model, or ``DP'' model, and we will use
the above expressions (Eqs.~\ref{apparent} and \ref{impedance1}) to
simulate experimental measurements.

\subsection{Numerical simulation of macroscopic measurements}
\label{sec-sim}

\subsubsection{Experiments of Gabriel et al. (1996)}

We first consider the experiments of Gabriel et
al.~\cite{Gabriel1996a,Gabriel1996,Gabriel1996c} who measured the
frequency dependence of electrical parameters for a large number of
biological tissues.  In these experiments, the biological tissue was
placed in between two capacitor plates, which were used to measure
the capacitance and leak current using the relation
$I_{\omega}=YV_{\omega}$, imposing the same current amplitude at all
frequencies.  Because the admittance value is proportional to
$\sigma_{\omega }^M +i\omega \epsilon_{\omega }^M$, measuring the
admittance provides direct information about $\sigma_{\omega }^M$ and
$\epsilon_{\omega }^M$.

To stay coherent with the formalism developed above, we will assume
that the capacitor has a spherical geometry. The exact geometry of
the capacitor should in principle have no influence on the frequency
dependence of the admittance, because the geometry will only affect
the proportionality constant between $\sigma_z$ and $Y_{\omega}$. In
the case of a spherical capacitor, by applying Eq.\ref{impedance1},
we obtain:
\begin{equation}
 Y_{\omega} = \frac{1}{R} + i\omega C = 4\pi\frac{R_1R_2}{R_2-R_1}[\sigma_{\omega}^M +i \omega \epsilon_{\omega}^M]=4\pi\frac{R_1R_2}{R_2-R_1}\sigma_z^M
\end{equation}

We also take into account the fact that the resistive part is
always greater than the reactive part for low frequencies
($<$~1000~Hz), which is expressed by
$$
 {\omega\epsilon_{\omega}^{M}} / {\sigma_{\omega}^{M}} << 1 ~ .
$$
This relation can be verified for example from the Gabriel et al.\
measurement~\cite{Gabriel1996}, where it is true for the whole 
frequency band investigated experimentally (between 10 and
10$^{10}$~Hz).

The real part of $\sigma_{\omega}^M=\sigma_z$ then takes the form
\begin{equation}
 \sigma_{\omega}^M \approx \frac{\sigma^M\sqrt{\omega}}{\sqrt{\omega}+k}
\cdot\frac{\omega\tau^2}{\omega\tau^2+1} 
\label{eq28}
\end{equation}
where $\tau = (\sqrt{\omega}+k_1) \tau_M$

By substituting this value of $\tau$, the inverse of the
conductivity (the resistivity) is given by: 
\begin{equation}
 \frac{1}{\sigma_{\omega}^M}\approx 
\frac{1}{\sigma^M}\cdot \left(1+\frac{k}{\sqrt{\omega}}\right)
\cdot\left(1+\frac{1}{\omega\tau_M^2 \left(\sqrt{\omega} + k_1\right)^2}\right)
= 
\frac{1}{\sigma^M}\cdot[1 +\frac{k}{\sqrt{\omega}} + \left(\frac{1}{\omega\tau_M^2}+\frac{k}{\omega^{3/2}\tau_M^2}\right)
\left(\frac{1}{\omega + 2k_1\sqrt{\omega}+k_1^2}\right)]
\label{sigmam}
\end{equation}

Finally, {to reproduce Gabriel et al.\ experiments, we assume
$k_1>>\sqrt{\omega}$.  By developing in series the last term (in
parenthesis) of Eq.~\ref{sigmam}, we obtain:}
\begin{equation}
\frac{1}{\sigma_{\omega}^M}\approx  \overline{K}_0 +\frac{\overline{K}_1}{\omega^{1/2}}
+\frac{\overline{K}_2}{\omega} +\frac{\overline{K}_3}{\omega^{3/2}}
=
K_0 +\frac{K_1}{f^{1/2}}
+\frac{K_2}{f} +\frac{K_3}{f^{3/2}}
\label{gab}
\end{equation}

\ \\ \centerline{------------------------ Figure~\ref{Puis1} here
------------------------}

Eq.~\ref{gab} corresponds to the conductivity $\sigma^M$, as
measured in the experimental conditions of Gabriel et al.\
experiments (the permittivity $\epsilon^M$ is obtained by applying
Kramers-Kronig relations).  Figure~\ref{Puis1} shows that this
expression for the conductivity can explains the measurements in
the frequency range of 10 to 1000~Hz, which are relevant for LFPs. 
To obtain this agreement, we had to assume in Eq.~\ref{apparent} a
relatively low Maxwell-Wagner time of the order of $0.15~s$ ($f_c =
1 / (2 \pi \tau_M)$ between 1~Hz and 10~Hz), $k_1 > \sqrt{\omega} >
k$ (for frequencies smaller than 100~Hz). 

{This value of Maxwell-Wagner time is necessary to explain
Gabriel's experiments, and may seem very large at first sight. 
However, the Maxwell-Wagner time is not limited by physical
constraints, because we have by definition
$\tau_{MW}=\frac{\epsilon_{\omega}}{\sigma_{\omega}}$.  In principle,
the value of $\sigma_{\omega}$ can be very small, approaching zero,
while $\epsilon_{\omega}$ can take very large values.  For example,
taking Gabriel et al.\ measurements in aqueous solutions of NaCl and
in gray matter~\cite{Gabriel1996}, gives values of $\tau_{MW}$
comprised between 1~ms and 100~ms for frequencies around 10~Hz.}

Thus, the model predicts that in the Gabriel et al.\ experiments,
the transformation of electric current carried by electrons to
ionic current in the biological medium necessarily implies an
accumulation of ions at the plates of the capacitor.  This ion
accumulation will in general depend on frequency, because the
conductivity and permittivity of the biological medium are
frequency dependent.  This will create a concentration gradient
across the biological medium, which will cause a ionic diffusion
current opposite to the electric current.  This ionic current will
allow a greater resulting current because surface polarization is
opposite to the electric field.  Figure~\ref{Puis1} shows that such
conditions give frequency-dependent macroscopic parameters
consistent with Gabriel et al.\ measurements.  

The parameter choices to obtain this agreement can be justified
qualitatively because the ionic diffusion constant on cellular
surfaces is probably much smaller than in the extracellular fluid,
such that $k_1>>k$.  This implies the existence of a frequency band
$B_f$ for which $\sqrt{\omega}$ is negligible with respect to
$k_1$, but not with respect to $k$ because these constants are
inversely proportional to the square root of their respective
diffusion coefficients.  Thus, the approximation that we suggest
here is that this band $B_f$ finishes around 100~Hz in the Gabriel
et al.\ experimental conditions.  It is important to note that this
parameter choice is entirely dependent on the ratio between ionic
diffusion current and the current produced by the electric field,
and thus will be depend on the particular experimental conditions.

It is interesting to note that the present model and the
phenomenological Cole-Cole model~\cite{Cole-Cole1941} predict
different behaviors of the conductivity for low frequencies
($<$10~Hz).  In the present model, the conductivity tends to zero
when frequency tends to zero, while in the Cole-Cole extrapolation,
it tends to a constant value~\cite{Gabriel1996a}.  The main
difference between these models is that the Cole-Cole model is
phenomenological and has never been deduced from physical
principles for low frequencies, unlike the present model which is
entirely deduced from well-defined physical phenomena.

\subsubsection{Logothetis et al. (2007) measurements}

We next consider the experiments of Logothetis et
al.~\cite{Logo2007}, which reported a resistive medium, in contrast
with Gabriel et al.\ experiments.  In these experiments, 4 electrodes
were aligned and spaced by 3~mm in monkey cortex.  The first and last
electrodes were used to inject current, while the two intermediate
electrodes were used to measure the extracellular voltage. 
{The voltage was measured at different frequencies and current
intensities.}

{One important point in this experimental setup is that the
intensity of the current was such that the voltage at the extreme
(injecting) electrodes saturates.  One of the consequences of this
saturation was to limit ionic diffusion effects, as discussed by the
authors~\cite{Logo2007}.}  This voltage saturation will diminish the
concentration difference near the source (we would have an
amplification if this was not the case).  It follows that, in the
Logothetis et al.\ experiments, the ratio between diffusion current
and electric field current is very small.  {Thus, in this case
we use values of parameters $k_i$ very different from those assumed
above to reproduce Gabriel et al.\ experiments, in particular $k_1 <<
\sqrt{\omega}$.}   As we will see in the Discussion section, this
situation may be different from the physiological conditions.

Nevertheless, the large distance between electrodes suggests that
the relation between current and voltage is linear because the
current density is roughly proportional to the inverse of squared
distance to the source.  Consequently, we can suppose that in the
Logothetis et al.\ experiments, the ionic gradient is negligible,
which prevents ionic diffusion currents.  Thus, in this experiment,
most -- if not all -- of the extracellular current is due to
electric-field effects.

In this situation, the conductivity (Eq.~\ref{apparent}) becomes:
\begin{equation}
\sigma_{\omega}^{M}\approx
\sigma^m
\cdot~\frac{(\omega\tau_M)^2}{1+(\omega\tau_M)^2} ~ ,
\end{equation}
which is similar to the ``Model P'' above.

Moreover, taking the same Maxwell-Wagner time $\tau_M$ as above 
for Gabriel et al.\ experiments (which corresponds to a cutoff 
frequency of 1~Hz), we have for frequencies above 10~Hz:
$$
\frac{(\omega\tau_M)^2}{1+(\omega\tau_M)^2} \approx 1 ~ ,
$$
similar to a resistive medium.

Thus, in the Logothetis et al.\ experiments, the saturation
phenomenon entrains current propagation in the biological medium as
if the medium was quasi-resistive for frequencies larger than
10~Hz.  This constitutes a possible explanation of the contrasting
results in Logothetis et al.\ and Gabriel et al.\ measurements.

\subsubsection{Frequency dependence of the power spectral density of
extracellular potentials}

The third type of experimental observation is the fact that the
power spectral density (PSD) of LFPs or EEG signals displays 1/f
frequency scaling~\cite{Bha2001,Bed2006a,Nov1997,Prit1992}.  To
examine if this 1/f scaling can be accounted for by the present
formalism, we consider a spherical current source embedded in a
continuous macroscopic medium.  We also assume that the PSD of the
current source is a Lorentzian, which could derive for example from
randomly occurring exponentially decaying postsynaptic
currents~\cite{Bed2006a} (see Fig.~\ref{DSP}).

To simulate this situation we used the ``diffusion-polarization''
model with ionic diffusion and electric field effects in a reactive
medium.  We have estimated above that surface polarization
phenomena have a cutoff frequency of the order of 1~Hz, and will
not play a role above that frequency.  So, if we focus on the PSD
of extracellular potentials in the frequency range larger than
1~Hz, we can consider only the effect of ionic diffusion (in
agreement with the Gabriel et al.\ experiments -- see above).  

Thus, we can approximate the conductivity as (see
Eq.~\ref{apparent}):
\begin{equation}
 \sigma_{\omega}^M = a \sqrt{\omega}
\end{equation}
where $a$ is a constant.

It follows that the extracellular voltage around a spherical current
source is given by (see Eq.~\ref{impedance}):
\begin{equation}
 V(r,\omega)= \frac{I_{\omega}}{4\pi a\sqrt{\omega}~r}=\frac{V(r,1)}{\sqrt{\omega}}
= \frac{V(R,1)R}{r\sqrt{\omega}}
\end{equation}
where $R$ is the radius of the source.

\ \\ \centerline{------------------------ Figure~\ref{DSP}
here ------------------------}

In other words, we can say that the extracellular potential is given
by the current source $I_\omega$ convolved with a filter in
1/$\sqrt{\omega~}$, which is essentially due to ionic diffusion
(Warburg impedance; see refs.~\cite{Diard1999,Hooge1997,Hooge1962}). 
A white noise current source will thus result in a PSD scaling as
1/f, and can explain the experimental observations, as shown in
Fig.~\ref{DSP}.  Experimentally recorded LFPs in cat parietal cortex
display LFPs with frequency scaling as 1/f for low frequencies, and
1/f$^3$ for high frequencies (Fig.~\ref{DSP}A-B).  Following the same
procedure as in ref.~\cite{Bed2006a}, we reconstructed the synaptic
current source from experimentally recorded spike trains
(Fig.~\ref{DSP}C-D).  The PSD of the current source scales as a
Lorentzian (Fig.~\ref{DSP}E) as expected from the exponential nature
of synaptic currents.  Calculating the LFP around the source and
taking into account ionic diffusion, gives a PSD with two frequency
bands, scaling in 1/f for low frequencies, and 1/f$^3$ for high
frequencies (Fig.~\ref{DSP}F).  This is the frequency scaling
observed experimentally for LFPs in awake cat cortex~\cite{Bed2006a}.
We conclude that ionic diffusion is a plausible physical cause of the
$1/f$ structure of LFPs for low frequencies.

{Two important points must be noted.  First, the
diffusion-plolarization model does not automatically predict $1/f$
scaling at low frequencies, but rather implements an $1/f$ filter,
which may result in frequency scaling with larger slopes.  Second,
the same experimental situation may result in different frequency
scaling, and this is also consistent with the diffusion-polarization
model.  These two points are illustrated in Fig.~\ref{DSP2} which
shows a similar analysis as Fig.~\ref{DSP} but during slow-wave sleep
in the same experiment.  The LFP is dominated by slow-wave activity
(Fig.~\ref{DSP2}A), and the different units display firing patterns
characterized by concerted ``pauses'' (gray lines in
Fig.~\ref{DSP2}B), characteristic of slow-wave sleep and which are
also visible in the reconstructed synaptic current
(Fig.~\ref{DSP2}C).  The PSD shows a similar scaling as $1/f^3$ as
for wakefulness, but the scaling at low frequencies is different
(slope around -2 at low frequencies; see Fig.~\ref{DSP2}D).  The PSD
reconstructed using the diffusion-polarization model displays similar
features (compare with Fig.~\ref{DSP2}E).  This analysis shows that
the diffusion-polarization model qualitatively accounts for different
regions of frequency scaling found experimentally in different
frequency bands and network states.}

\ \\ {\centerline{------------------------ Figure~\ref{DSP2}
here ------------------------}}

\subsection{Measurement of frequency dependence}
\label{sec-test}

In this final section, we examine a possible way to test the model
experimentally.  The main prediction of the model is that, in
natural conditions, the extracellular current perpendicular to the
source is dominated by ionic diffusion.  The experiments realized
so far~\cite{Gabriel1996a,Gabriel1996,Gabriel1996c,Logo2007} used
macroscopic currents that did not necessarily respect the correct
current flow in the tissue.  

We suggest to create more naturalistic current sources by creating
ionic currents using a micropipette placed in the extracellular
medium.  By using periodic current injection during very short time
$\Delta t$ compared to the period (small duty cycle), we can
measure using the same electrode the extracellular voltage $V_{pr}$
(using a fixed reference far away from the source).  If the period
of the source is shorter than the relaxation time of the system,
the voltage will integrate, which is due to charge accumulation.  

Because $V_{pr}$ is directly proportional to the amount of charge
emitted as a function of time during $\Delta t$ (capacitive effect
of the extracellular medium), the time variation of $V_{pr}$ is
directly proportional to the ionic diffusion current.  In such
conditions, if the extracellular medium is purely resistive, as
predicted by Logothetis et al.\ experiments, the relaxation time
should be very small, of the order of $10^{-12}~s$ \cite{Bed2006},
which would prevent any integration phenomena and charge
accumulation for frequencies below $10^{12}$~Hz.  If the medium has
a slower relaxation, due to polarization and ionic diffusion, then
we should observe voltage integration and charge accumulation for
physiological frequencies ($<$~1000~Hz).

To illustrate the difference between these two situations, we
consider the simplest case of a non-reactive medium (as in Model~D
above), in which the current can be produced by ionic diffusion or
by the electric field, or by both.  To calculate the time variations 
of ionic concentration and extracellular voltage, we consider the 
current density:
\begin{equation}
\vec{j} = D\nabla e[c] + \sigma \vec{E}
\end{equation}
According to the differential law for charge conservation and
Poisson law, we have:
\begin{equation}
\nabla \vec{j} + \frac{\partial \rho}{\partial t}=D~\nabla^2\rho ~+ 
~\frac{\sigma }{\epsilon }~\rho~ + ~\frac{\partial \rho}{\partial t} =0
\label{pe}
\end{equation}
When ionic diffusion is negligible compared to Ohm's law, we have:
$$
\frac{\partial \rho}{\partial t}+\frac{\sigma }{\epsilon }\rho  \approx 0
$$
It follows that the charge density is given by:
\begin{equation}
 \rho = \rho_o~ \exp\left(-\frac{\sigma}{\epsilon}t\right)
\label{sol1}
\end{equation}

On the other hand, if ionic diffusion is the primary cause of current
propagation in the extracellular medium, then the relaxation time
should be much larger and thus, integration should be observed.  When
ionic diffusion is dominant, we have:
$$
  \frac{\partial \rho}{\partial t}+ D~\nabla^2 \rho \approx 0
$$
instead of Eq.~\ref{pe}.

The general solution is:
\begin{equation}
\rho = \frac{1}{\sqrt{2Dt}}
~\int_{0}^{\infty} \rho(r,0)e^{-\frac{-r^2}{4Dt}}dr
\label{sol2}
\end{equation}

The difference between the expressions above (Eqs.~\ref{sol1} and
\ref{sol2}) shows that the time variation of charge density is
different according to which current dominates, electric field
current or ionic diffusion current.  The same applies to the
electric potential between the electrode and a given reference,
because this potential is linked to charge density through
Poisson's law.  Therefore, this experiment would be crucial to
clearly show which of the two current primes for currents
perpendicular to the source (this would not apply to longitudinal
currents, like axial currents in dendrites).  In the hypothetical
case of dominant ionic diffusion, the cortex would be similar to a
Warburg impedance and one can estimate the macroscopic diffusion
coefficient using Eq.~\ref{sol2}.

Thus, using a micropipette injecting periodic current pulses, it
should be possible to test the capacity of the medium to create
charge accumulation for physiological frequencies.  If this is the
case, this would constitute evidence that ionic currents are
non-negligible in the physiological situation.


\section{Discussion}

In the present paper, we have proposed a framework to model local
field potentials, and which synthesizes previous measurements and
models.  This framework integrates microscopic measurements of
electric parameters (conductivity $\sigma$ and permittivity
$\epsilon$) of extracellular fluids, with macroscopic measurements of
those parameters ($\sigma_{\omega}^M$, $\epsilon_{\omega}^M$) in
cortical tissue~\cite{Gabriel1996,Logo2007}.  It also integrates
previous models of LFPs, such as the {\it continuum
model}~\cite{Bed2004}, which was based on a continuum hypothesis of
electric parameters variations in extracellular space, or the {\it
polarization model}~\cite{Bed2006}, which explicitly considered
different media (fluid and membranes) and their polarization by the
current sources.  The present model is more general and also
integrates ionic diffusion, which is predicted as a major determinant
of the frequency dependence of LFPs.  This ``diffusion-polarization''
model also accounts for observations of $1/f$ frequency scaling of
LFP power spectra, which is due here to ionic diffusion, and is
therefore predicted to be a consequences of the genesis of the LFP
signal, rather than being {solely} due to neuronal activity (see
ref.~\cite{Bed2006a}).  {Finally, this work suggests that
ephatic interactions between neurons can occur not only through
electric fields but that ionic diffusion should also be considered
in such interactions.} 

As discussed in Section~\ref{simplif}, the present model rests on
several approximations, which were necessary to obtain the analytic
expressions used here.  These approximations were that current
sources were considered as monopolar entities (longitudinal currents
such as axial currents in dendrites were not taken into account), the
current source was spherical and the extracellular medium was
considered isotropic.  Because multipolar effects can be
reconstructed from the superposition of monopoles~\cite{Purc1984},
the monopolar configuration should not affect the results on
frequency dependence, as long as the extracellular current
perpendicular to the source is considered.  {Similarly, the
exact geometry of the current source should have no influence on the
frequency dependence far away from the sources.  However, in the
immediate vicinity of the sources, the geometric nature and the
synchrony of synaptic currents can have influences on the power
spectrum~\cite{Pettersen2008}.  Another assumption is that the
extracellular medium is isotropic, which was justified within the
macroscopic framework followed here.}  These factors, however, will
influence the exact shape of the LFPs.  More quantitative models
including a more sophisticated geometry of current sources and the
presence of membrane excitability and action potentials should be
considered (e.g., see refs.~\cite{Gold2006,Pettersen2008}).

The main prediction of the present model is that ionic diffusion is
an essential physical cause for the frequency dependence of LFPs.  We
have shown that the presence of ionic diffusion allows the model to
account quantitatively for the macroscopic measurements of the
frequency dependence of electric parameters in cortical
tissue~\cite{Gabriel1996}.  Ionic diffusion is responsible for a
frequency dependence of the impedance as $1/\sqrt{\omega~}$ for low
frequencies ($<$~1000~Hz), which directly accounts for the observed
$1/f$ frequency scaling of LFP {and EEG power spectra during
wakefulness}~\cite{Bha2001,Bed2006a,Nov1997,Prit1992} (see
Fig.~\ref{DSP}).  {Note that the EEG is more complex because it
depends on the diffusion of electric signals across fluids, dura
matter, skull, muscles and skin.  However, this ``filtering'' is of
low-pass type, and may not affect the low-frequency band, so there is
a possibility that the 1/f scaling of EEG and LFPs have a common
origin.  The present model is consistent with the view that this
apparent ``$1/f$ noise'' in brain signals is not generated by
self-organized features of brain activity, but is rather a
consequence of the genesis of the signal and its propagation through
extracellular space~\cite{Bed2006a}.}

{It is important to note that the fact that ionic diffusion may
be responsible for $1/f$ frequency scaling of LFPs is not
inconsistent with other factors which may also influence frequency
scaling.  For example, the statistics of network activity -- and more
generally network state -- can affect frequency scaling.  This is
apparent when comparing awake and slow-wave sleep LFP recordings in
the same experiment, showing that the $1/f$ scaling is only seen in
wakefulness but $1/f^2$ scaling is rather seen during
sleep~\cite{Bed2006a} (see Fig.~\ref{DSP2}).  In agreement with this,
recent results indicate that the correlation structure of synaptic
activity may influence frequency scaling at the level of the membrane
potential, and that correlated network states scale with larger (more
negative) exponents~\cite{Marre2007}.}

We also investigated ways to explain the measurements of Logothetis
et al.~\cite{Logo2007}, who reported that the extracellular medium
was resistive and therefore did not display frequency dependence, in
contradiction with Gabriel et al.~\cite{Gabriel1996} measurements. 
{We summarize and discuss our conclusions below.}

{In Gabriel et al.\ experiments~\cite{Gabriel1996}, one measures
permittivity and conductivity in the medium in between two metal
plates.  This forms a capacitor, which (macroscopic) complex
impedance is measured.  This measure actually consists of two
independent measurements, the real and imaginary part of the
impedance.  These values are used to deduce the macroscopic
permittivity and macroscopic conductivity of the medium.  However, at
the interface between the medium and the metal plates, the flow of
electrons in the metal corresponds to a flow of charges in the
tissue, and a variety of phenomena can occur, which can interfere the
measurement.  The accumulation of charges that occurs at the
interface between the electrode and the extracellular fluid implies a
polarization impedance, which depends on the interaction between ions
and the metal plate.  Because this accumulation of charge implies a
variation of concentration, the flow of ions may involve an important
component of ionic diffusion.} 

{In Logothetis et al.\ experiments, a system of 4 electrodes is
used, the two extreme electrodes inject current in the medium, while
the two electrodes in the middle are exclusively used to measure the
voltage.  This system is supposed to be more accurate than Gabriel's,
because the electrodes that measure voltage are not subject to charge
accumulation.  However, the drawback of this method are nonlinear
effects.  The magnitude of the injected current is such that the
voltage at the extreme electrodes saturates.  This voltage saturation
also implies saturation of concentration (capacitive effect between
electrodes), which limits ionic diffusion currents. Thus, the ratio
between ionic diffusion currents and the currents due to the electric
field is greatly diminished relative to Gabriel et al.\ experiments.}

{We think that natural current sources are closer to Gabriel et
al.\ situation for several reasons.  First, the magnitude of the
currents produced by biological sources is far too low for saturation
effects.  Second, the flow of charges across ion channels will
produce perturbations of ionic concentration, which will be
re-equilibrated by diffusion.  The effects may not be as strong as
the perturbations of concentrations induced by Gabriel et al.\ type
of experiments, but ionic diffusion should play a role in both cases.
This is precisely one of the aspects that should be evaluated in
further experiments.} 

{The Logothetis et al.\ experiments were done using a
4-electrode setup which neutralizes the influence of electrode
impedance on voltage measurements~\cite{Geddes1997,McAdams1992}. 
This system was used to perform high-precision impedance
measurements, also avoiding ionic diffusion effects~\cite{Logo2007}. 
Indeed, these experimental conditions, and the apparent resistive
medium, could be reproduced by the present model if ionic diffusion
was neglected.  The present model therefore formulates the strong
prediction that ionic diffusion is important, and that any
measurement technique should allow ionic diffusion to reveal the
correct frequency-dependent properties of impedance and electric
parameters in biological tissue.}

The critical question that remains to be solved is whether, in
physiological conditions, ionic diffusion plays a role as important
as suggested here.  We propose a simple method to test this
hypothesis.  The frequency dependence could be evaluated by using an
extracellular electrode injecting current in conditions as close as
possible to physiological conditions (a micropipette would be
appropriate).  By measuring the integration of the extracellular
voltage following periodic current injection, one could estimate the
``relaxation time'' of the medium with respect to charge
accumulation.  If this relaxation time occurs at time scales relevant
to neuronal currents (milliseconds) rather than the fast relaxation
predicted by a purely resistive medium (picoseconds), then ionic
diffusion will necessarily occur in physiological conditions, which
would provide evidence in favor of the present mechanism.


\begin{appendix}

\section*{Appendices}

\section{Estimation of ionic diffusion current vs.\ electric
diffusion in sea water and in cortex.\label{annexeB}}

{In this appendix, we evaluate the ratio between ionic diffusion
currents and electric field currents in the extracellular space
directly adjacent to the source.  This ratio measures if the ionic
diffusion current perpendicular to the membrane is greater than the
electric field current.  We will design this ratio by the term
$r_{ie}$.}

We have in general
\begin{equation}
\vec{j}_{Total} = eD\nabla[C] +\sigma \nabla V ~ ,
\end{equation}
where the first term in right hand is the electric current density
produced by ionic diffusion, while the second term is that produced
by differential Ohm's law.

For a displacement $\Delta r =10~ nm$ in the direction across the 
membrane (from inside to outside), we have approximatively: 
\begin{equation}
 \vec{j}_{Total}\cdot d\vec{r}\simeq\vec{j}_{Total}\cdot\vec{\Delta r}
= e\Delta[C]_{\Delta r = 10 nm}+\sigma_{\omega} \Delta V_{\Delta r = 10 nm} 
\end{equation}

Suppose that we have a spherical cell of $10~\mu m$ radius, embedded
in sea water and at resting potential.  The resting membrane
potential is a dynamic equilibrium between inflow and outflow of
charges, in which these two fluxes are equal on (temporal) average. 
Fluctuations of current around this average in the extracellular
medium around the membrane have all characteristics of thermal
noise~\cite{Nyquist1928} because the shot noise (see
refs.~\cite{Buck1985} and \cite{Vas1983}) is zero when the current is
zero on average, such that the net charge on the external side of the
membrane varies around a mean value with the same characteristics as
white noise (thermal noise).  These fluctuations will therefore be
present also at the level of the membrane potential.  In this
appendix, we evaluate the order of magnitude of the electric current
caused by ionic diffusion, relative to the electric field for this
situation of dynamic equilibrium.

First, the ratio between the membrane voltage noise and the variation
of total charge concentration is given by
\begin{equation}
\Delta Q_{tot} = C\Delta V_{membrane}=k_1\Delta V_{membrane}
=1.25\times 10^{-11}\Delta V_{membrane} ~ ,
\end{equation}
because the cell's capacitance is given by $C=4\pi R^2C_m=0.04\pi
R^2$ ($C_m \simeq10^{-2}~F/m^2$) 

Second, mass conservation imposes:
\begin{equation}
\Delta Q_{tot} = e*v_{eff}*\Delta[C]_{tot} ~ ,
\end{equation}
where $e=1.69\times 10^{-19}~C$ and $v_{eff}$ is the volume of the
spherical shell containing the charges.  Because the charges are not
uniformly distributed inside the cell, but rather distributed within
a thin spherical shell adjacent to the membrane, because the
electric field developed across the membrane is very intense (of the
order of $\frac{70 \times 10^{-3}}{7 \times 10^{-9}}~V/m=10^7~V/m$),
Thus, the width of the
shell is of the order of $dR\simeq 10^{-4}R< 1~ nm$, where the volume
of the spherical shell is approximately equal to $4\pi R^2dR$.  In
this case, we have for monovalent ions ($|z|=1$)
\begin{equation}
\Delta Q_{tot}  = k_2 \Delta[C]_{tot} \simeq
2.2\times10^{-38}\Delta[C]_{tot}=2.2\times10^{-38}\Delta[C]_{\Delta r
= 10 nm}
\end{equation}
if we assume that the variation of concentration on the adjacent
border of the exterior surface of the cell is over a width of 
$10~ nm$. 

In this case, we have
\begin{equation}
\frac{\Delta [C]_{\Delta r = 10 nm}}{\Delta  V_{membrane}} =\frac{k_1}{k_2} \simeq 10^{27}~ C/m^3V
\end{equation}

Third, the potential difference between the cell surface and 
$10~ nm$ away of it, is given by
\begin{equation}
\Delta V_{\Delta r =10 nm} = \Delta V_{membrane} -\frac{R\Delta V_{membrane}}{R+r}\simeq 10^{-3}\Delta V_{membrane}
\end{equation}

Consequently, the ratio between ionic diffusion current and electric
diffusion current caused by thermal noise in sea water obeys
\begin{equation}
r_{ie}\approx\frac{eD_{sea}\Delta [C]_{\Delta r = 10 nm}\Delta
t}{\sigma_{\omega}^{sea}\Delta V_{\Delta r =10 nm}
}=\frac{e D_{sea}k_1}{\sigma_{\omega}^{sea} k_2}\simeq
\frac{10^{2}}{\sigma_{\omega}^{sea}} ~ ,
\end{equation}
where the diffusion constant of $K^+$ or $Na^+$ in sea water is of
the order of $10^{-9}~\frac{m^2}{s}$.  This implies that the ratio
$r$ is much larger than 1 for frequencies $<$1000~Hz because
$\sigma_{\omega}^{sea}$ of sea water is necessarily $<2~S/m$.

{Because tortuosity is given by $\lambda=\sqrt{D_{sea} /
D_{cortex}}$, and is comprised between 1.6 and 2.2 (for small and
large molecules, respectively) in cerebral
cortex~\cite{Nicho1998,Nicho2005,Rus1998}, the macroscopic diffusion
constant in cortex is certainly larger than $D_{sea}/10$. Thus, we
have
\begin{equation}
r_{ie}^{cortex} > \frac{10}{\sigma_{\omega}^{cortex}}
\label{rapM}
\end{equation}
where $\sigma_{f = 100 ~Hz}^{cortex}\simeq 0.1~S/m$ 
(see ref.~\cite{Gabriel1996c}).
}

This evaluation shows that the phenomenon of ionic diffusion is
essential to determine the current field in the cortex.

Finally, we note that we did not need to evaluate the absolute
magnitude of $\Delta V$ in our evaluation.  This evaluation is valid
for a physical situation where we have a permanent white noise over a
distance of 10~nm, independently of the intensity of this noise (which
in practice will depend on many factors, such as the size of the
cell, the number of ion channels, etc).

\section{Frequency scaling of ionic diffusion}
\label{secscal}

In this appendix, we calculate the frequency dependence of ionic
diffusion current outside of a spherical current source.  We consider
a constant variation of ionic concentration, $\Delta X_i$, on the
surface of the source, and a null variation at an infinite distance
(Warburg conditions).

The diffusion equation for a given ionic species is
\begin{equation}
\frac{\partial \Delta X_i}{\partial t} = D_{i}\nabla^2 \Delta X_i ~ ,
\end{equation}
where $\Delta X_i$ is the perturbation of concentration $X_i$ of ion
$i$ around the steady-state value, and $D_i$ is the associated
diffusion coefficient.  This diffusion coefficient depends of the
ionic species considered and the structure of the medium.

Because the geometry of the problem and the limit conditions respect
spherical symmetry, we use spherical coordinates.  In this
coordinate system, we have
\begin{equation}
\frac{\partial \Delta X_i}{\partial t}=D_i[\frac{\partial^2 \Delta X_i}{\partial r^2} + 
\frac{2}{r}\frac{\partial \Delta X_i}{\partial r} ] 
\end{equation} 
because $\Delta X_i$ does not depend on $\theta$ and of $\Phi$
(spheric symmetry).  

The Fourier transform of $\Delta X_i$ with respect to time gives:
\begin{equation}\label{Laplace}
\frac{\partial^2 X_{i_{\omega}}}{\partial r^2}+\frac{2}{r}\frac{\partial \Delta X_{i_{\omega}}}{\partial r} 
= \frac{d^2 X_{i_{\omega}}}{d r^2}+\frac{2}{r}\frac{d \Delta X_{i_{\omega}}}{d r}
=\frac{i\omega}{D_i} \Delta X_{i_{\omega}}
\end{equation}
which general solution is given by:
\begin{equation}
 \Delta X_{i_{\omega}} = A(\omega )\frac{ e^{\sqrt{\frac{i\omega }{D_i}}~r}}{r}
+B(\omega )\frac{ e^{-\sqrt{\frac{i\omega }{D_i}}~r}}{r}
\end{equation}

For a variation of concentration at the source border which is
independent of frequency and which satisfies the Warburg hypothesis
(the variation of concentration tends to zero at an infinite
distance~\cite{Diard1999,Tay1995}), we have:
\begin{equation}
 \Delta X_{i_{\omega}}(r) = 
 \Delta X_{i_{\omega}}(R)\cdot \frac{R~e^{-\sqrt{\frac{i\omega }{D_i}}~(r-R)}}{r}
\end{equation}
where $r$ is the distance between the center of the source and 
$R$ is the radius of the source.

Thus, the electric current density produced by ionic diffusion is
given by:
\begin{equation}
 \vec{j_i}(r)= ZeD_i\frac{\partial \Delta X_i}{\partial r}\hat{r}  
=-ZeD_i~(~\frac{1}{r}+\sqrt{\frac{i\omega~}{D_i}}~)~\Delta X_{i_{\omega}}(r)~\hat{r}
\end{equation}
where $Ze$ is the charge of ions $i$.  

Because we can consider that the source and extracellular medium
form a spherical capacitor, the voltage difference between the
surface of the source and infinite distance is given by
 $ZeC\Delta X_{i_{\omega}}(R)$ where $C$ is the capacitance value.
Thus, the electric impedance of the medium is given by:
\begin{equation}
 Z_{\omega}=\frac{C}{D_i(\frac{1}{R}+\sqrt{\frac{i\omega~}{D_i}})}
\end{equation}

For a source of radius $R = 10~\mu m$ and a macroscopic ionic
diffusion coefficient of the order of $10^{-11}~ m^2/s$, and for
frequencies $>$ 1~Hz, we can approximate the impedance by:
\begin{equation}
 Z_{\omega} \approx \frac{C}{\sqrt{i\omega D_i~}}
\end{equation}

The same expression for the impedance is also obtained in cylindric
coordinates or planar Cartesian coordinates (not shown). 

{Note that if several ionic species are present, then the
superposition principle applies (Fick equations are linear) and
therefore the contribution of each ion will add-up.  The diffusion
constants for different ions are of the same order of magnitude (for
Na$^+$, K$^+$, Cl$^-$, Ca$^{2+}$), so no particular ion would be
expected to dominate.}

\end{appendix}


\subsection*{Acknowledgments}

{We thank Christoph Kayser, Nikos Logothetis, Axel Oeltermann
and anonymous reviewers for useful comments on the manuscript.}
Research supported by CNRS, ANR and the European Community (FACETS
project).



\label{Bibliographie-Fin}

\clearpage
\section*{Tables}

\begin{table}[ht]
$$
\begin{array}{c|ccc}
n ~~~& \Delta\epsilon_n & \tau_n ~(s) & \alpha_n \\
\hline
1 ~~~& 4.50\times 10^{1} ~&~ 7.96 \times 10^{-12} ~& ~0.10 \\
2 ~~~& 4.00\times 10^{2} ~&~15.92 \times 10^{-9} ~& ~0.15 \\
3 ~~~& 2.0\times 10^{5} ~& ~106.1 \times 10^{-6} ~& ~0.22 \\
4 ~~~& 4.5\times 10^{7} ~&~5.305 \times 10^{-3} ~& ~0.00 \\
\end{array}
$$
\caption{Parameter values for the parametric model of Gabriel et
al.~\cite{Gabriel1996} (see Eq.~\ref{Eq4}).}

\end{table}


\clearpage
\section*{Figure Legends}

\begin{figure}[bht]

\caption{{Illustration of the two main physical phenomena
involved in the genesis of local field potentials.  A given current
source produces an electric field, which will tend to polarize the
charged membranes around the source, as schematized on the top.  The
flow of ions across the membrane of the source will also involve
ionic diffusion to re-equilibrate the concentrations.  This diffusion
of ions will also be responsible for inducing currents in
extracellular space.  These two phenomena influence the frequency
filtering and the genesis of LFP signals, as explored in this paper.}}

\label{polariz}
\end{figure}

\begin{figure}[bht]

\caption{Monopole and dipole arrangements of current sources.  A. 
Scheme of the extracellular medium containing a quasi-dipole (gray)
representing a pyramidal neuron, with soma and apical dendrite
arranged vertically.  B. Illustration of one of the monopoles of the
dipole. The extracellular space is represented by cellular processes
of various size (circles) embedded in a conductive fluid.  The dashed
lines represent equipotential surfaces.  The $\widehat{ab}$ line
illustrates the fact that the extracellular fluid is linearly connex.}

\label{ModeleSimple} 
\end{figure}

\begin{figure}[bht]

\caption{Models of macroscopic extracellular conductivity compared to
experimental measurements in cerebral cortex.  The experimental data
(labeled ``G'') show the real part of the conductivity measured in
cortical tissue by the experiments of Gabriel et
al.~\cite{Gabriel1996}.  The curve labeled ``E'' represents the
macroscopic conductivity calculated according to the effects of
electric field in a non-reactive medium.  The curve labeled ``D'' is
the macroscopic conductivity due to ionic diffusion in a non-reactive
medium.  The curve labeled ``P'' shows the macroscopic conductivity
calculated from a reactive medium with electric-field effects
(polarization phenomena).  The curve labeled ``DP'' shows the
macroscopic conductivity in the full model, combining the effects of
electric polarization and ionic diffusion.  Every model was fit to
the experimental data by using a least-square procedure, and the best
fit is shown.  The DP model's conductivity is given by
$\frac{1}{\sigma_{\omega }^M} = K_0 + \frac{K_1}{f^{1/2}}
+\frac{K_2}{f }+\frac{K_3}{f^{3/2}}$, with $K_0=10.84 $, $K_1=-19.29
$, $K_2 =180.35 $ and $K_3=52.56 $.  The experimental data (G) is the
parametric Cole-Cole model~\cite{Cole-Cole1941} which was fit to the
experimental measurements of Gabriel et al.~\cite{Gabriel1996}.  This
fit is in agreement with experimental measurements for frequencies
larger than 10~Hz.  No experimental measurements exist for
frequencies lower than 10~Hz, and the different curves show different
predictions from the phenomenological model of Cole-Cole and the
present models. }

\label{Puis1}
\end{figure}

\begin{figure}[bht]

\caption{Simulation of 1/f frequency scaling of LFPs {during
wakefulness}.  A. LFP recording in the parietal cortex of an awake
cat.  B. Power spectral density (PSD) of the LFP in log scale,
showing two different scaling regions with a slope of -1 and -3,
respectively.  C. Raster of 8 simultaneously-recorded neurons in the
same experiment as in A.  D.  Synaptic current calculated by
convolving the spike trains in C with exponentials (decay time
constant of 10~ms).  E. PSD calculated from the synaptic current,
shown two scaling regions of slope 0 and -2, respectively.  F. PSD
calculated using a model including ionic diffusion (see text for
details).  The scaling regions are of slope -1 and -3, respectively,
as in the experiments in B. Experimental data taken from
ref.~\cite{Des1999}; see also ref.~\cite{Bed2006a} for details of
the analysis in B-D.}

\label{DSP}
\end{figure}

\begin{figure}[bht]

\caption{{Simulation of more complex frequency scaling of LFPs
during slow-wave sleep.  A. Similar LFP recording as in
Fig.~\ref{DSP}A (same experiment), but during slow-wave sleep.  B. 
Raster of 8 simultaneously-recorded neurons in the same experiment as
in A.  The vertical gray lines indicate concerted pauses of firing
which presumably occur during the ``down'' states.  C. Synaptic
current calculated by convolving the spike trains in B with
exponentials (decay time constant of 10~ms).  D. Power spectral
density (PSD) of the LFP in log scale, showing the same scaling
regions with a slope of -3 at high frequencies as in wakefulness (the
PSD in wake is shown in gray in the background).  At low frequencies,
the scaling was close to $1/f^2$ (gray line; the dotted line shows
the $1/f$ scaling of wakefulness).  E. PSD calculated from the
synaptic current in C, using a model including ionic diffusion.  This
PSD reproduces the scaling regions of slope -2 and -3, respectively
(gray lines).  The low-frequency region, which was scaling as $1/f$
in wakefulness (dotted lines), had a slope close to -2.  Experimental
data taken from ref.~\cite{Des1999}.}}

\label{DSP2}
\end{figure}


\clearpage
\begin{figure} 
\centering
 \includegraphics[width=16cm]{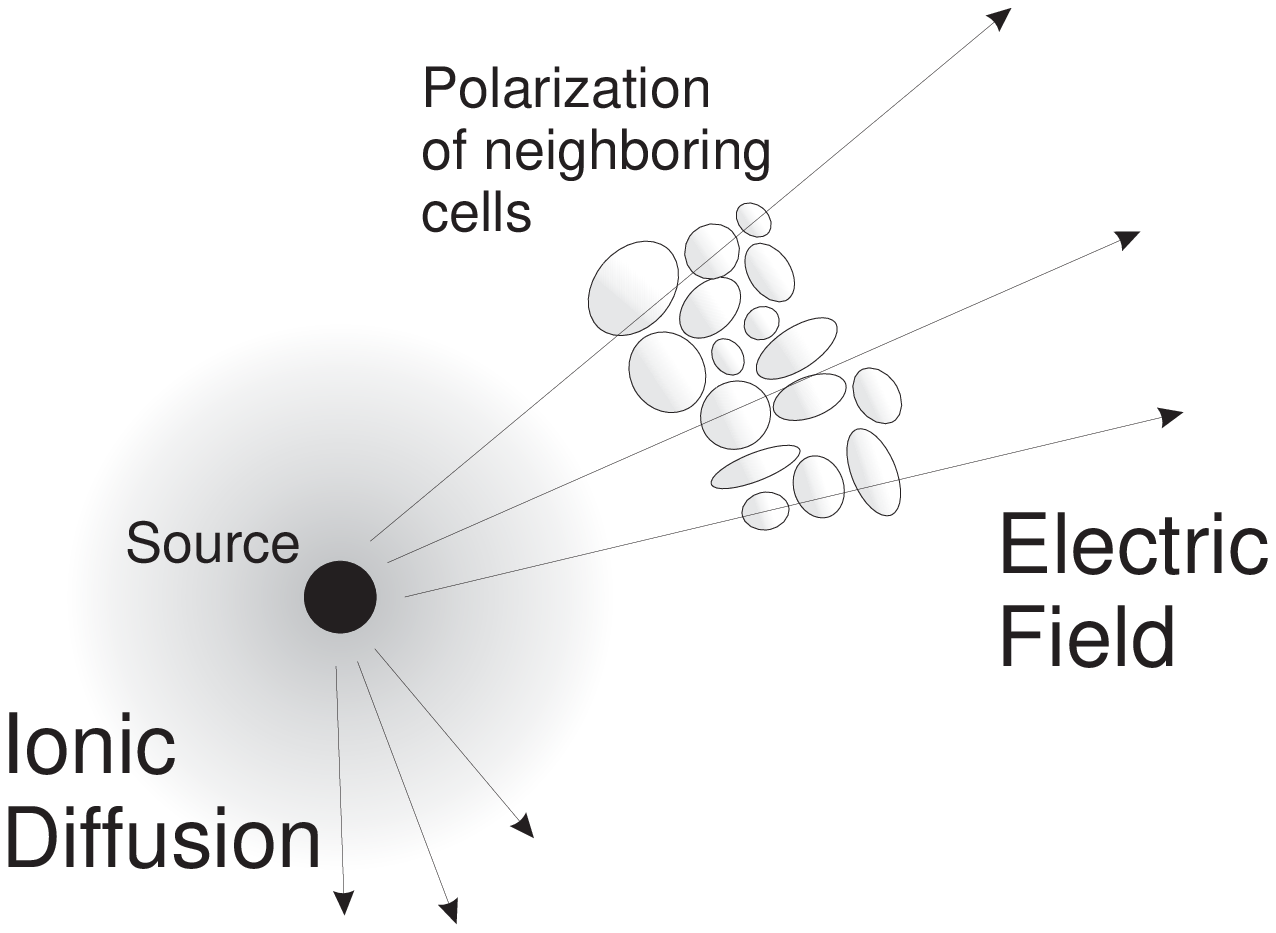}

\ \\ Figure~\ref{polariz}

\end{figure} 

\clearpage
\begin{figure} 
\centering
 \includegraphics[width=12cm]{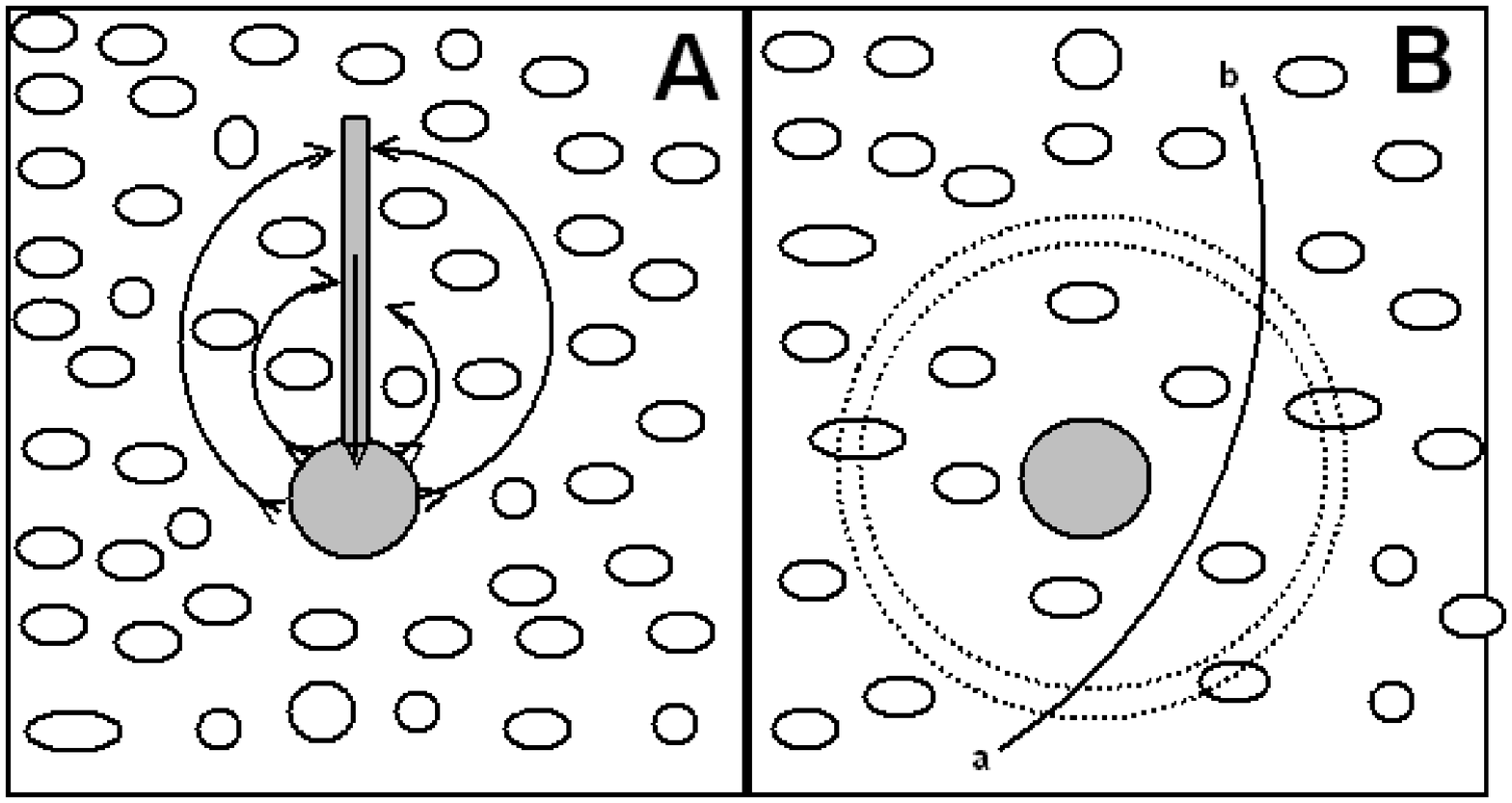}

\ \\ Figure~\ref{ModeleSimple}

\end{figure} 

\clearpage

\begin{figure} 
\centering
 \includegraphics[width=12cm]{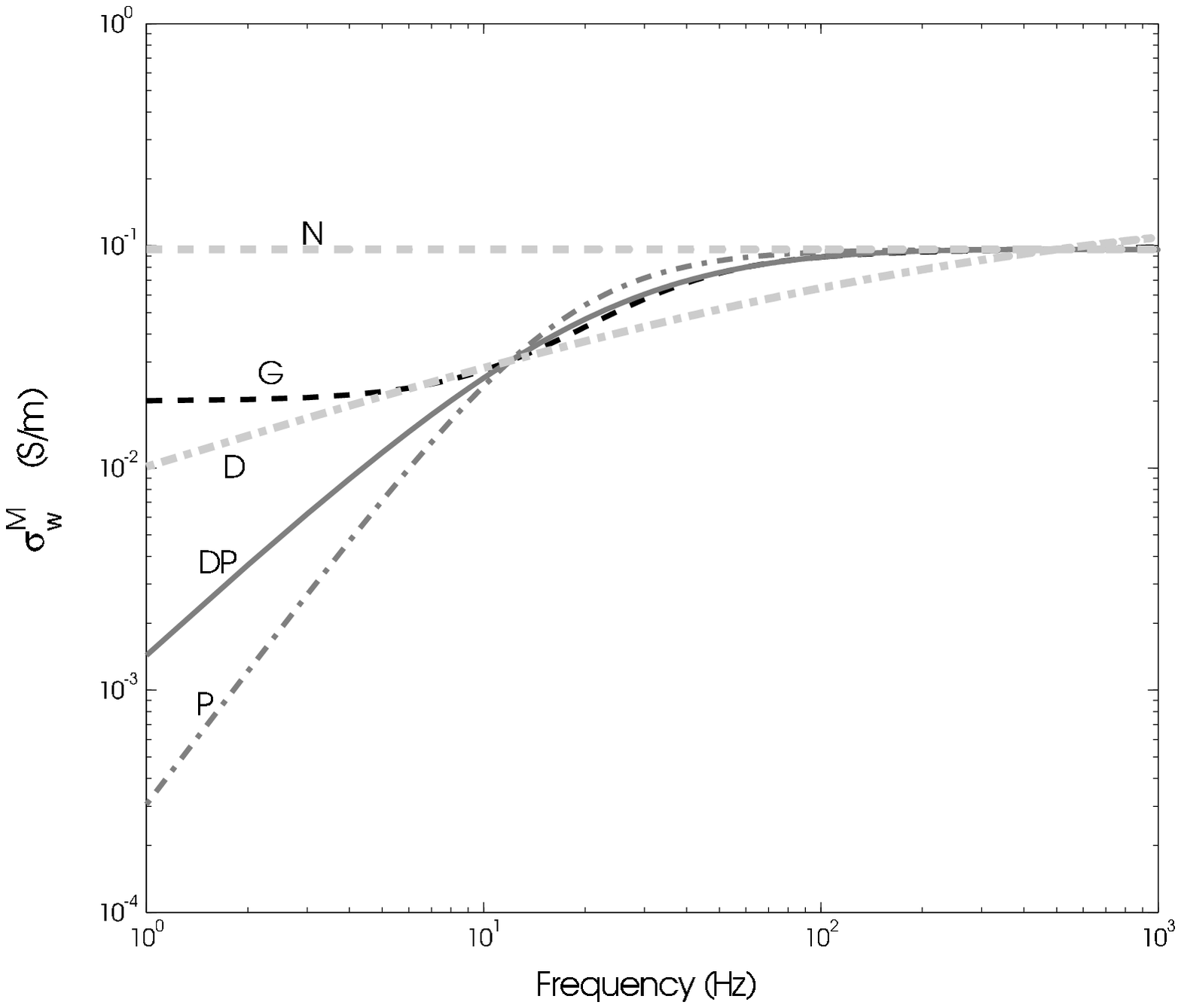}

\ \\ Figure~\ref{Puis1}

\end{figure} 

\clearpage

\begin{figure} 
\centering
 \includegraphics[width=16cm]{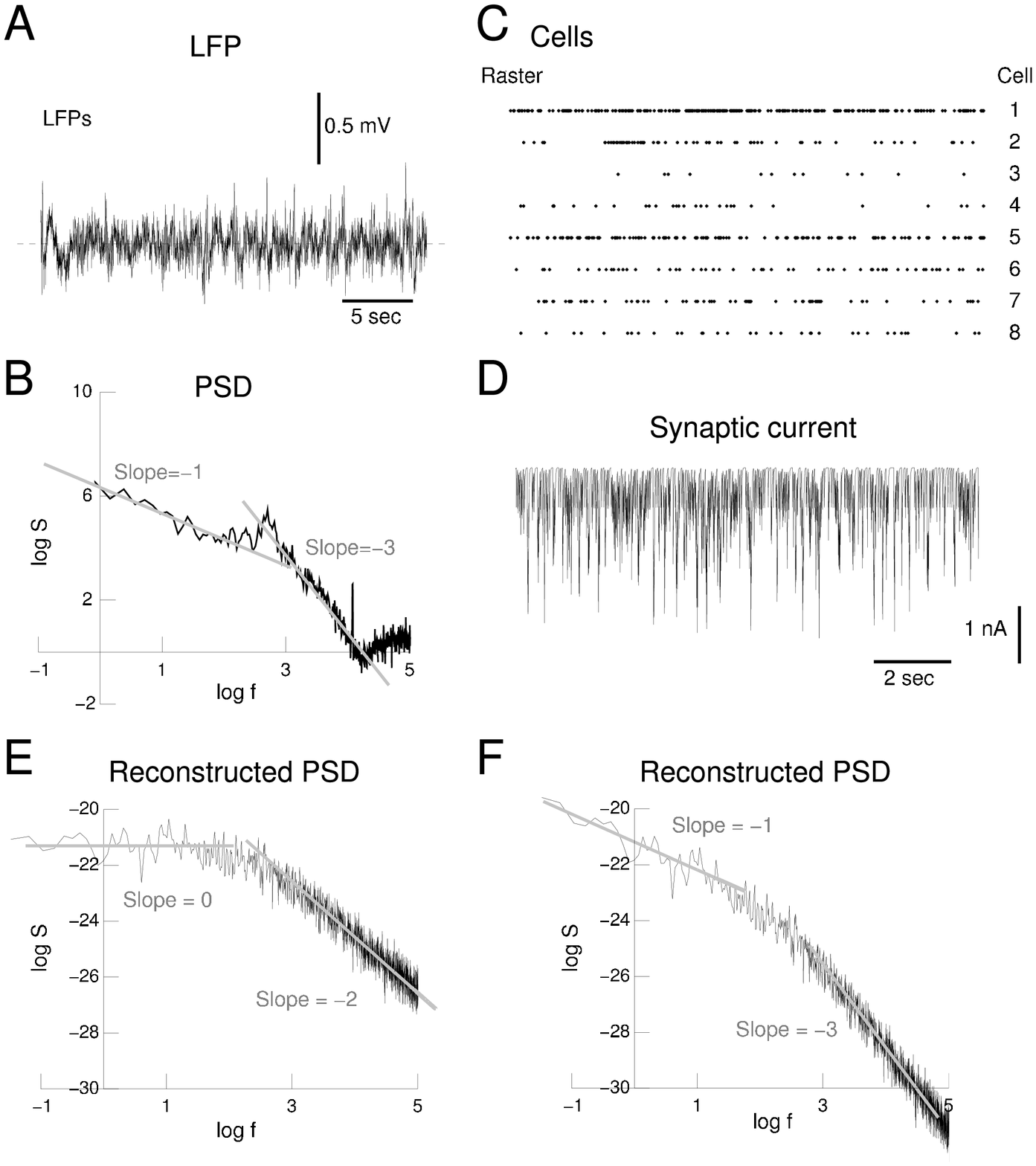}

\ \\ Figure~\ref{DSP}

\end{figure} 

\clearpage

\begin{figure} 
\centering
 \includegraphics[width=13cm]{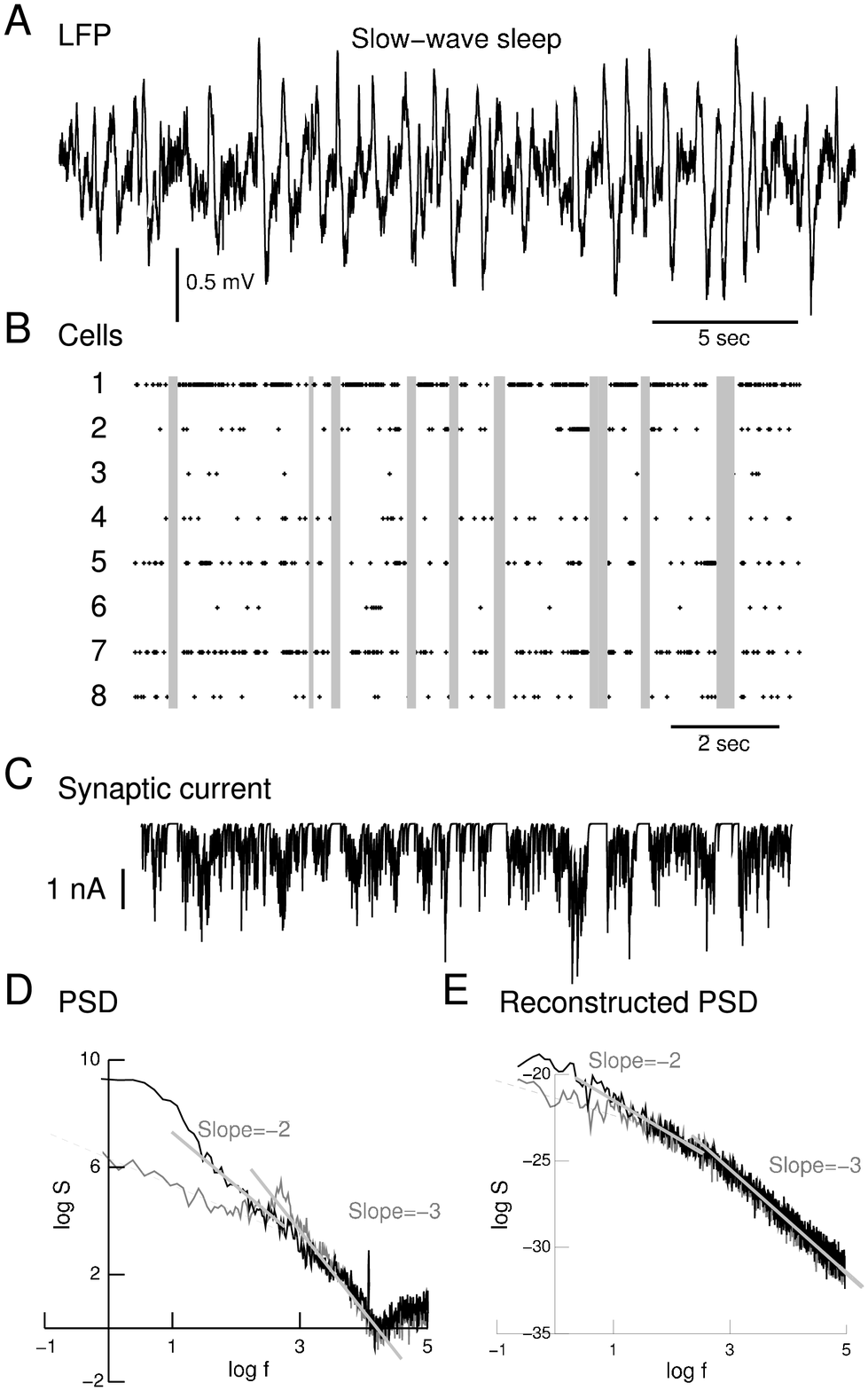}

\ \\ Figure~\ref{DSP2}

\end{figure} 

\end{document}